\documentclass[namedreferences]{solarphysics}

\usepackage[hyperref,optionalrh,solaromanenum]{spr-sola-addons} 
\usepackage{graphicx}                    
\usepackage{color}                       
\usepackage{breakurl}                         

\newcommand{\cc}{{cm$^{-3}$}}

\begin{document}

\begin{article}

\begin{opening}

\title{Anticipating Solar Flares}

%
 \author[addressref={1,2,3},corref,email={hugh.hudson@glasgow.ac.uk}]{\inits{}\fnm{Hugh}~\lnm{Hudson}\orcid{0000-0001-5685-1283}}

%
\runningauthor{Hudson}
\runningtitle{Anticipating Solar Flares}

\address[id={1}]{SSL, UC Berkeley, CA USA}
\address[id={2}]{School of Physics and Astronomy, University of Glasgow, UK}
\address[id={3}]{Western Kentucky University, Bowling Green KY USA}

\begin{abstract}
Solar flares commonly have a ``hot onset precursor event'' (HOPE), detectable from soft X-ray observations.
This requires subtraction of pre-flare fluxes from the non-flaring Sun prior to the event, fitting an isothermal emission model to the flare excess fluxes by comparing the GOES passbands at 1--8~\AA\ and 0.5--4~\AA, and plotting the timewise evolution of the flare emission in a diagram of temperature \textit{vs} emission measure.
The HOPE then appears as an initial ``horizontal branch'' in this diagram.
It precedes the non-thermal impulsive phase of the flare and thus the flare peak in soft X-rays as well.
We use this property to define a ``flare anticipation index'' (FAI), which can serve as an alert for observational programs aimed at solar flares based on near-real-time soft X-ray observations.
This FAI gives lead times of a few minutes and produces very few false positive alerts even for flare brightenings too weak to merit NOAA classification.
\end{abstract}

%

\end{opening}

%
\section{Introduction}\label{sec:intro} 

Solar flares have precursor signatures of several types, which may appear in coronal and chromospheric observations.
Perhaps the most remarkable of these consist of the pre-flare activations of filaments, which  may then erupt; for example, the \textit{Skylab} astronauts famously used real-time monitoring of H$\alpha$ images to identify the beginnings of eruptive flares, now known to identify with coronal mass ejections (CMEs).
In soft X-rays, the precursor activity may show up as a characteristic slow increase ramping up to the ``impulsive phase'' characterizing strongly non-thermal phenomena such as particle acceleration and the energization of CMEs \citep{1970ApJ...162.1003K} and thus the full development of flare emissions.
The subject of pre-flare activity, and of soft X-ray precursor, has an extensive literature, which this article does not attempt to review.
We do note \cite{2023A&A...671A..73P}, who have used machine learning to follow several observables, and have thereby recognized clear chromospheric patterns doubtless related to the GOES coronal phenomenon described below as HOPE (``Hot Onset Precursor Event'').


Recently \cite{2021MNRAS.501.1273H} have examined the soft X-ray precursors, finding them to have characteristic properties and suggesting universality for the process.
Further work has strengthened this conclusion \citep{2023MNRAS.tmp.2291D,2023A&A...679A.139B}.
The HOPE phenomenon therefore has the potential to become a tool for a flare alert on few-minute time scales.
In this paper we describe a ``flare anticipation index'' (FAI) based on the standard GOES soft X-ray observations, with the objective of anticipating flare occurrence far enough ahead in time to enable campaign-style observational programs a sufficient warning for observations aimed at impulsive-phase physics.
We find the FAI to be extremely reliable, even for events too weak to be classified in the standard ABCMX spectrum of NOAA flare reports.
This finding strongly confirms the case for universality of the HOPE phenomenon, even though we do not yet understand the physics behind it. 

In practical terms, the GOES soft X-rays provide a convenient basis for an FAI, given the vast database available and the near-real-time (latency of a few minutes) data currently provided by NOAA.
This GOES-based approach, though successfully anticipating soft X-ray events well below NOAA's C-class, could lead to other FAI methods that may have better still better sensitivity).
Note that flare anticipation is not the same thing as flare forecasting.
The HOPE appears to be just the earliest recognizable feature of a flare, for which actual prediction remains a difficult problem.

This article studies the GOES FAI based on a single 3-day sample of the real-time NOAA X-ray database, which have one-minute cadence (2024-01-02T12:00 through 2024-01-05T12:00).
Such a quick sample suffices to justify operational deployment of this very simple algorithm for assisting with observing campaigns aimed at flare/CME origins.
The Appendix describes how observers can implement an FAI alert. 
The rest of this paper explains more about how it works, but the main purpose here is just to show how to use it.

\section{Identifying the soft X-ray HOPE}\label{sec:hope} 

\begin{figure}[htbp]
\centering
    \includegraphics[width=\textwidth]{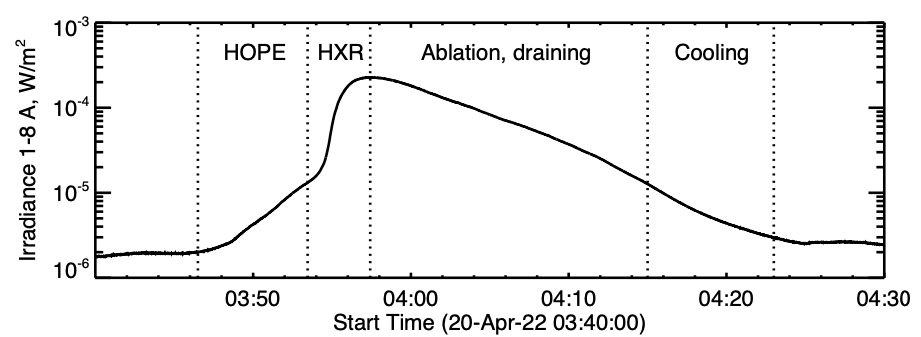}
       \caption{\textit{The soft X-ray developmental phases of the X-class flare SOL2022-04-20, showing the prominent HOPE preflare increase.
       See the text and Figure~\ref{fig:jakimiec} for further explanation.}
        }  
        \label{fig:fai_plot}
\end{figure}

The X-class flare SOL2022-04-20 illustrates the HOPE pattern extremely well.
Figure~\ref{fig:fai_plot} shows the time development in the GOES/XRS soft 1-8~/AA/ channel.
The ramp-up at the outset is the HOPE phase, as labeled; the impulsive phase (labeled HXR) shows when hard X-rays and other nonthermal effects happen, including ablation (``evaporation''); this merges into arcade development and draining, and eventually simple cooling and associated field shrinkage.
This particular event displays these phases with unusual clarity.

\begin{figure}[htbp]
\centering
    \includegraphics[width=\textwidth]{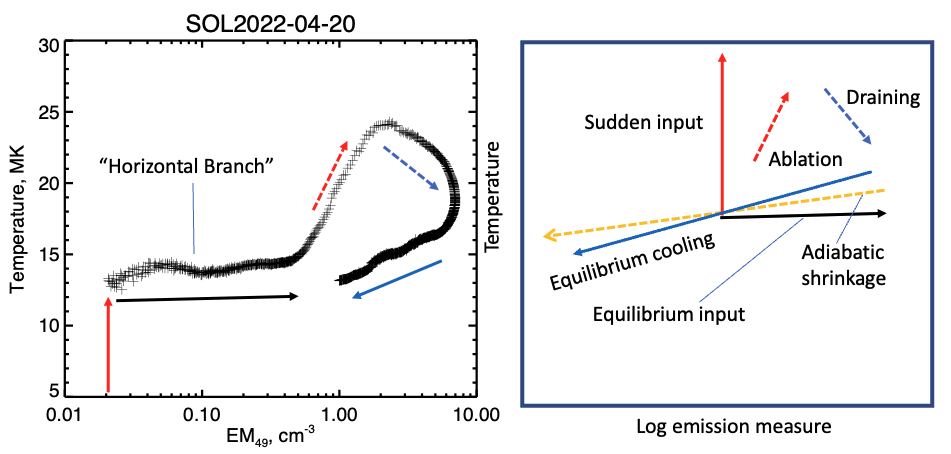}
       \caption{\textit{Left, a representative [EM, T] diagnostic displaying the hot onset effect.
       Right, descriptions of the directions followed by the diagnostic point.
       The striking and useful characteristic feature of the HOPE phenonenon is the ``horizontal branch,'' during which the emission measure grows steadily while the temperature remains approximately constant.
       These flare data for SOL2022-04-20 (X2.2) show the constituent parts of the diagnostic diagram exceptionally clearly, but this pattern is normally present.}
        }  
        \label{fig:jakimiec}
\end{figure}

We explain the time development with a [T, EM] diagnostic diagram related to that described by \cite{1986AdSpR...6f.237J}, who phrased it in terms of theory as [T, $\sqrt{n_e}$] rather than directly in terms of the observables.
Figure~\ref{fig:jakimiec} shows a clear example of such a diagram, along with a sketch relating its features to the flare development.
The [T, EM] variables come from a direct isothermal fit, via SolarSoft tools \citep{1998SoPh..182..497F}, to the two-channel flux residuals above estimated background levels.
The newly recognized feature underpinning the FAI development is the initial horizontal branch, during which emission measure grows steadily, sometimes punctuated by microflares, while the temperature of the increasing mass remains roughly constant.
The analysis in this article confirms the universality of the soft X-ray horizontal branch as a requirement for flare development.

In the explanatory sketch (right panel of Figure~\ref{fig:jakimiec}) the red arrows show the result of energy input into coronal plasma.
The impulsive phase, which in the example shown begins at $EM \approx 0.02 \times 10^{49}$~cm$^{-3}$, initiates the ablation of large amounts of new hot material as ``evaporation.'' 
This motion of the correlation point describes a clockwise loop of the [T,EM] trajectory.
In contrast to this, the initial point already exhibits a  temperature well above that of the quiet Sun or an active region. 
This does not represent ``heating'' in the sense of temperature increase.

\section{The GOES-based Flare Anticipation Index (FAI)}\label{sec:fai} 

We can use the appearance of the initial horizontal branch of the [EM,T] diagram by screening on the values of the timewise motion of the isothermal fits, [dEM/dT, T].
A first guess at such an FAI used timewise differencing on the GOES near-realtime data.
The FAI algorithm requires 5 parameters, and Table~1 lists first-guess default values.
The unit EM$_{49}$ for the volumetric emission measure is the SolarSoft standard value of 10$^{49}$~\cc.

\begin{table}[h]
\label{tab:fai_params}
\caption{GOES FAI Parameters}
\begin{tabular}{lll}     
\hline                     
Parameter & Default & Significance \\
  \hline
Integration time & 1 min & Set by GOES quicklook data\\
Difference time $\Delta t$ & 5 min & Initial guess\\
EM increment& 0.005  EM$_{49}$ & Explored in this article\\
Temperature range & [7,14]~MK  & Explored in this article\\
FAI duration & 3~min  & Not explored in this article\\
\hline
\end{tabular}
\end{table}

\begin{figure}[htbp]
\centering
    \includegraphics[width=0.9\textwidth]{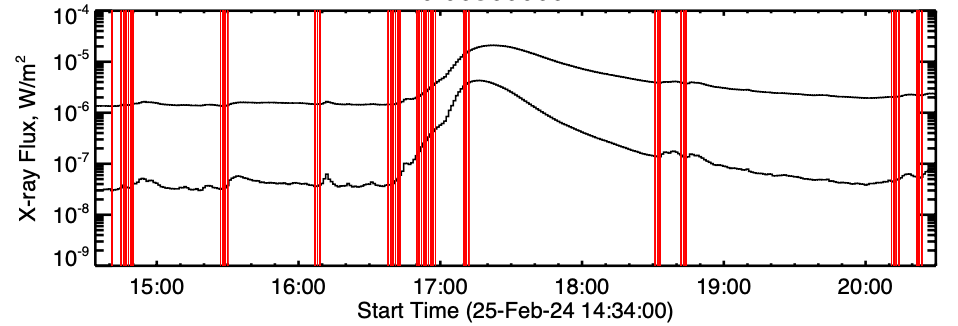}
       \caption{\textit{A quick check of the default parameters for a six-hour stretch of real-time GOES data. The red vertical lines show minutes for which the flag was set for these values.
       As can be seen, there were {\bf no significant} false positives {\bf but some apparent} false negatives {\bf (see text)}.
       The FAI run with a lower incremental EM parameter readily detects the weak A-class events missed at the default setting. }
        }  
        \label{fig:fai_realtime}
\end{figure}

For a randomly chosen six-hour interval of GOES real-time data, the FAI algorithm -- as operated with the default parameters -- generated Figure~\ref{fig:fai_realtime}.
The flare anticipation worked extremely well, with 100\% true positives (all flags preceded GOES 1-8~\AA\ maxima) and at times of a few minutes prior to the flares' impulsive phases.
Note one small anomaly, however; in the precursor to the major event (SOL2024-02-25T17:22, M2.1) the EM increased monotonically as the horizontal branch of the [EM,T] diagnostic evolved, but with some irregularity.
The gaps in the flag sequence in the major flare in this example turned out to result from the default setting of the high-temperature limit; adjusting it upwards resolved this issue.
The description in the Appendix suggests [6,20]~MK.

This article studies a 3-day quicklook data set in detail: 2024-01-02T12:00  through 2024-01-05T12:00.
During this time interval NOAA reported 20 flares, ranging from class C1.2 to M3.8, a period with a relatively high soft X-ray background level.
We list these in Table~2 along with the correspondingt FAI ``anticipation times'' (GOES 1-8~\AA\ peak minus time of first FAI flag, with values 13.4 $\pm\ 6.0$~min).

\begin{table}[h]
\label{tab:GOES_times}
\caption{GOES Time Comparisons}
\begin{tabular}{llllrr}     
\hline 
Date & Start & Peak & End & Class & Anticipation\\
&&&&&(minutes)\\
\hline 
 2-JAN-24&  13:42& 13:45& 13:49&  C1.2 &   6  \\                
 3-JAN-24&  02:54& 02:59& 03:05&  C1.2 &   6  \\                            
 3-JAN-24&  10:00& 10:10& 10:14&  C1.8 &  16 \\                      
 3-JAN-24&  10:14& 10:18& 10:24&  C3.0 &   8  \\                    
 3-JAN-24&  13:29& 13:34& 13:44&  C1.3 &   8  \\                             
 3-JAN-24&  14:42& 14:57& 15:15&  C1.8 &  24 \\                                 
 3-JAN-24&  16:13& 16:21& 16:28&  C1.3 &  13 \\                               
 3-JAN-24&  16:49& 16:56& 17:08&  C1.5 &  10 \\                             
 4-JAN-24&  00:13& 00:25& 00:51&  C2.1 &  15  \\                              
 4-JAN-24&  01:08& 01:16& 01:22&  M1.1 &  11  \\                             
 4-JAN-24&  01:22& 01:55& 02:12&  M3.8 &  27  \\                                
 4-JAN-24&  07:19& 07:28& 07:48&  C1.5 &  11  \\                             
 4-JAN-24&  08:55& 09:06& 09:16&  C1.7 &  14 \\                              
 4-JAN-24&  09:16& 09:36& 09:42&  C3.0 &  15 \\                                
 4-JAN-24&  10:20& 10:30& 10:35&  C2.2 &  12 \\                                 
 4-JAN-24&  17:20& 17:31& 17:39&  C3.3 &  12 \\                               
 5-JAN-24&  00:32& 00:52& 01:08&  C3.2 &  15 \\                               
 5-JAN-24&  02:43& 02:51& 02:56&  C1.7 &   9  \\                             
 5-JAN-24&  04:18& 04:25& 04:30&  C1.6 &  11 \\                             
 5-JAN-24&  07:55& 08:09& 08:19&  C3.7 &  24\\ 
\hline
\end{tabular}
\end{table}

\begin{figure}[htbp]
\centering
    \includegraphics[width=0.44\textwidth]{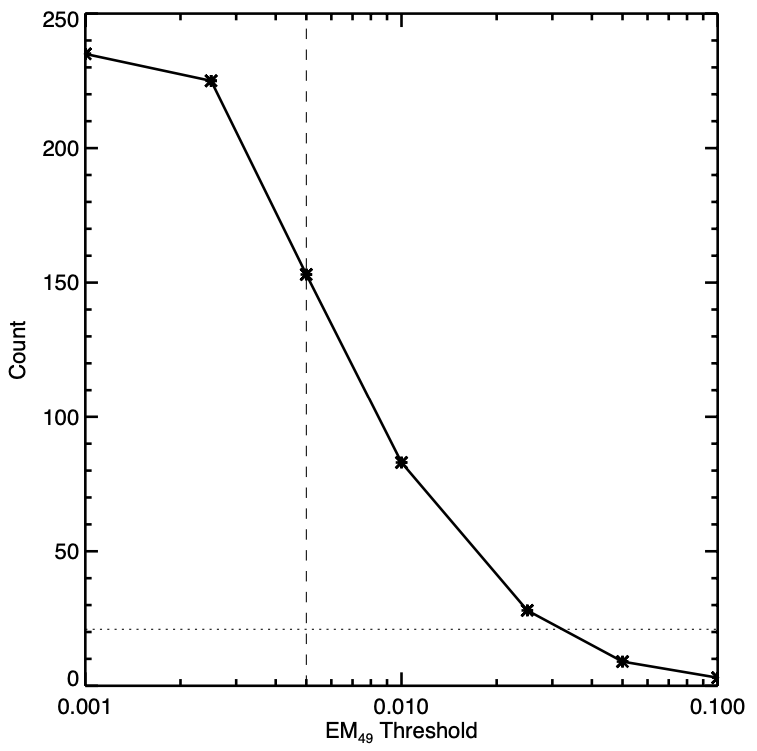}
   \includegraphics[width=0.52\textwidth]{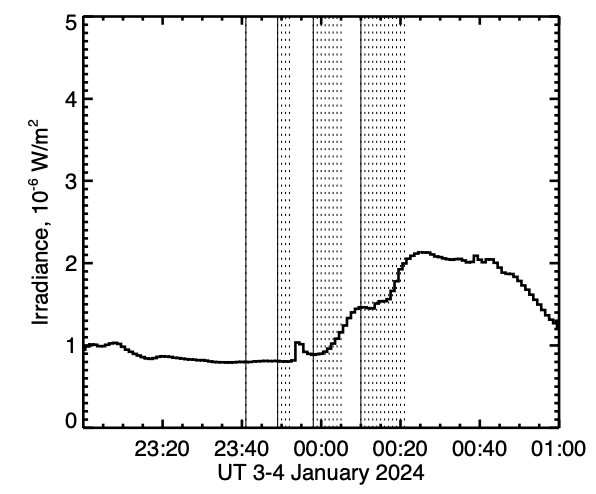}
       \caption{\textit{Left, numbers of events as a function of the EM increment parameter.
       The dotted line shows the SolarSoft flare count for the interval, and the vertical dashed
       line shows the default value for EM increment.
       Right, a section of time series showing the FAI failure for the energetically A8-class event SOL2024-01-04T00:39 (see text), plotted on a linear scale with C1 units 
       (10$^{-6}$~W/m$^2$).
       This plot used an EM increment of 0.001 EM$_{49}$; the solid lines show the first of each set of consecutive flags.}
        }  
        \label{fig:on_stats}
\end{figure}

The default values for the five parameters listed in Table~1 worked well for the test interval.
We have explored adjusting the EM increment and show the resulting event counts in Figure~\ref{fig:on_stats} (left panel).
At the default value the flare numbers greatly exceed the SolarSoft count of 20, which corresponds to an EM increment of about $0.03 \times 10^{49}$~\cc.
Reducing the FAI threshold for EM increment below its default value reliably returns many more events than the NOAA classification recognizes.

The physical interpretation of the EM increment is the growth of emission measure per five minutes, basically a rate of change.
Some flares, particularly slow and/or powerful, meet this criterion for many consecutive 1-minute samples.
This indicates a monotonic increase in total EM at roughly constant temperature.
The counts in Figure~\ref{fig:on_stats} correspond to the first sample of any such grouping, thus marking the earliest anticipation time for a given flare.

The right panel of Figure~\ref{fig:on_stats} illustrates how the FAI may fail for the weakest
events, showing particularly a tiny C2-class peak (background-subtracted to an A8 level), SOL2024-01-04, not forewarned even at an emission-measure criterion of $EM_{49} = 0.001$ (or lower).
This reflects the practical limit of the FAI algorithm, considering the digital steps of the GOES telemetry and the sensors' background fluctuations as well as the background level.

\section{Parameter dependences (FAI)}\label{sec:parms} 

\begin{figure}[htbp]
\centering
    \includegraphics[width=0.9\textwidth]{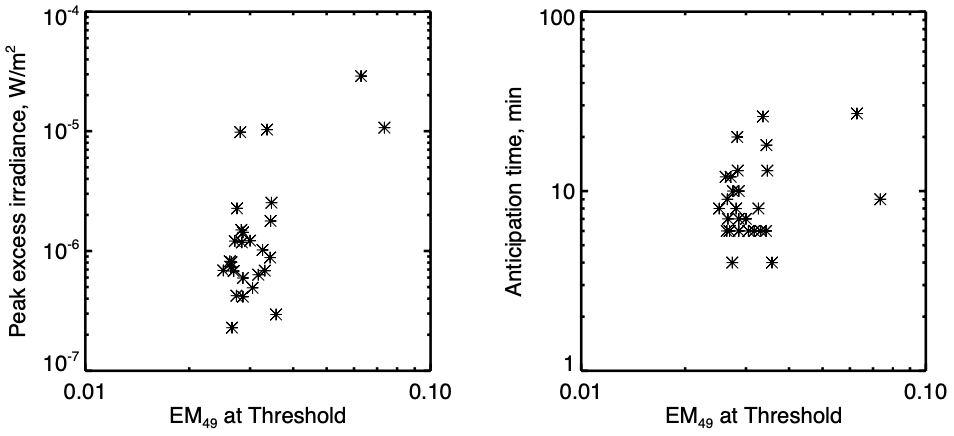}
   \includegraphics[width=0.9\textwidth]{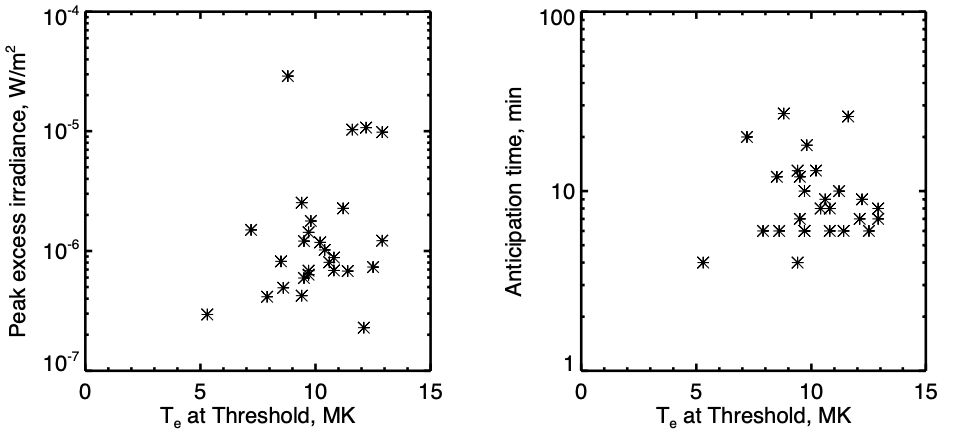}
       \caption{\textit{Correlations between FAI parameters (for trigger EM upper panels, and T$_e$ lower panels) and observables (left panel for flare peak flux, and right panel for anticipation time).}
        }  
        \label{fig:correlations}
\end{figure}

Do the FAI parameters not only anticipate flare occurrence, but also correlate with its peak flux or other properties?
This short sample does not yield decisive results here, but the decades of archival GOES data will allow future studies with much greater precision.
For the 3-day quicklook data set discussed in this article there is already a hint of peak-flux prediction (Figure~\ref{fig:correlations}).

The M-class flares appear in the upper right corner of the upper left panel of Figure~\ref{fig:correlations}, consistent with a predictive capability for flare magnitude. 
This correlation is well borne out anecdotally by the author in reference to high M-class and X-class flares not studied in detail here. 
Also anecdotally, there may be an onset temperature correlation for major events, in that the default temperature range [7,14]~MK needed to be increased to [6,20]~MK to set FAI flags for the X-class flare SOL2024-02-09.
The recipe detailed in the Appendix uses [6,20]~MK for the temperature range.

A future more complete study will need to map probabilities between the set of five model parameters and  additional items of practical interest: how long must we wait for the flare, and what will its duration be? We already have established a correlation between emission-measure increment and flare magnitude.

\section{Conclusions}\label{sec:concl} 

This article has briefly described a flare anticipation index (FAI) based on the standard GOES two-channel solar soft X-ray fluxes.
The purpose here is a practical one, aiming at helping solar observers to predict imminent flare occurrence.
The sample studied here is a 3-day realtime dataset, and the FAI worked extremely well, with 100\% true positives and no false negatives at the default parameter set.
Campaign-style observations benefiting from a few minutes' warning of flare occurrence can therefore use the algorithm as-is.
The limitations of the FAI include the latency of the GOES realtime data, about 4 min, and the whole-Sun nature of the data.
A different database solving these problems will doubtless enable a successor for the present FAI; ideally the new data would consist of soft X-ray imaging observations.

This data set hints at a predictive capability of the FAI: greater emission-measure increments in the HOPE phase correspond to more energetic flares.
As noted, this predictive capability may also extend to the HOPE T$_e$ values, but a systematic study of archival data will be needed to quantify this.
The GOES FAI used here runs on IDL under SolarSoft \citep{1998SoPh..182..497F} and the working script can be obtained from the author.\footnote{hugh.hudson@glasgow.ac.uk}

Finally, a brief comment on the significance of the uniform FAI success.
The steady growth of emission measure prior to a flare can now be recognized as a universal property of solar flares on all magnitude scales. 
This slow development itself constitutes the initial unstable action of the flaring plasma, culminating in the main flare instability itself, with the impulsive phase, particle acceleration, evaporation, and perhaps the CME launch.
We note that for weak events at GOES A and B class, the FAI anticipates the flare occurrence but often finds a higher onset value of GOES T$_e$ than during the flare that eventually develops.

\begin{acks}
The author thanks the University of Glasgow for hospitality, and Sarah Paterson for reading the draft manuscript prior to submission.
Andrew Jones and Tom Woods have also provided valuable discussion.
\end{acks}

\section{Appendix: Implementation}
This recipe generates a useful FAI from one-minute samples of near-real-time GOES soft X-ray data:
\begin{enumerate}
\item Download near-realtime GOES data at \url{https://services.swpc.noaa.gov/json/goes/primary/xrays-6-hour.json}
\item Extract the two broadband data values, XRS\_A and XRS\_B, the fluxes the 0.5--4~\AA\ and 1--8~\AA\ bands, respectively.
\item Form running differences between the latest value and the value $\Delta t$~min prior, to create two new timeseries.
\item Interpret these running-difference flux values in terms of isothermal temperature and emission measure $[T_6, EM_{49}]$ via SolarSoft {\sc goes\_tem.pro} or equivalent.
These are estimates of the electron temperature in units of 10$^6$~K, and volumetric emission measure in units of 10$^{49}$~cm$^{-3}$.
\item Set a flare flag where $T_a < T_6 < T_b$ and $EM_{49} > Y$, where $[T_a, T_b,\mathrm{and}\ Y]$ are free parameters. 
Nominal values of the parameters for $\Delta t = 5$~min set at [6, 20, 0.1] detect an M-class flare about 15~min prior to GOES maximum in XRS\_B, with no false positives or false negatives.
\end{enumerate} 

Adjustment of the $Y$ parameter detects flares at different magnitudes, and the user can work out an appropriate interface for interpreting the near-real-time FAI signature.
At the time of writing, the GOES input data from the link above have an unavoidable 4-5~min latency.

\clearpage
\bibliographystyle{spr-mp-sola}
\bibliography{fai.bib}  

\begin{thebibliography}{7}
\ifx\bisbn     \undefined \def\bisbn  #1{ISBN #1}\fi
\ifx\binits    \undefined \def\binits#1{#1}\fi
\ifx\bauthor   \undefined \def\bauthor#1{#1}\fi
\ifx\batitle   \undefined \def\batitle#1{#1}\fi
\ifx\bjtitle   \undefined \def\bjtitle#1{\textit{#1}}\fi
\ifx\bvolume   \undefined \def\bvolume#1{\textbf{#1}}\fi
\ifx\byear     \undefined \def\byear#1{#1}\fi
\ifx\bissue    \undefined \def\bissue#1{#1}\fi
\ifx\bfpage    \undefined \def\bfpage#1{#1}\fi
\ifx\blpage    \undefined \def\blpage #1{#1}\fi
\ifx\burl      \undefined \def\burl#1{#1}\fi
\ifx\href      \undefined \def\href#1#2{#2}\fi
\ifx\betal     \undefined \def\betal{et al.}\fi
\ifx\bctitle   \undefined \def\bctitle#1{#1}\fi
\ifx\beditor   \undefined \def\beditor#1{#1}\fi
\ifx\bbtitle   \undefined \def\bbtitle#1{\textit{#1}}\fi
\ifx\bedition  \undefined \def\bedition#1{#1}\fi
\ifx\bseriesno \undefined \def\bseriesno#1{\textbf{#1}}\fi
\ifx\blocation \undefined \def\blocation#1{#1}\fi
\ifx\bsertitle \undefined \def\bsertitle#1{\textit{#1}}\fi
\ifx\bsnm      \undefined \def\bsnm#1{#1}\fi
\ifx\bsuffix   \undefined \def\bsuffix#1{#1}\fi
\ifx\bparticle \undefined \def\bparticle#1{#1}\fi
\ifx\barticle  \undefined \def\barticle#1{}\fi
\ifx\binstitute  \undefined \def\binstitute#1{#1}\fi
\ifx\bpublisher  \undefined \def\bpublisher#1{#1}\fi
\ifx\doiurl    \undefined \def\doiurl#1{\href{#1}{DOI}}\fi
\makeatletter
\def\safeHref#1#2#3{\in@{http}{#2}\ifin@\href{#2}{#3}\else\href{#1#2}{#3}\fi}
\makeatother
\ifx\adsurl    \undefined
  \def\adsurl#1{\safeHref{https://ui.adsabs.harvard.edu/abs/}{#1}{ADS}}\fi
\ifx\arxivurl  \undefined
  \def\arxivurl#1{\safeHref{http://arxiv.org/abs/}{#1}{arXiv}}\fi
\ifx\botherref \undefined \def\botherref#1{}\fi
\ifx\url       \undefined \def\url#1{#1}\fi
\ifx\bchapter  \undefined \def\bchapter#1{}\fi
\ifx\bbook     \undefined \def\bbook#1{}\fi
\ifx\bcomment  \undefined \def\bcomment#1{#1}\fi
\ifx\oauthor   \undefined \def\oauthor#1{#1}\fi
\ifx\citeauthoryear \undefined\def \citeauthoryear#1{#1}\fi
\def\endbibitem {}
\ifx\bconflocation  \undefined \def\bconflocation#1{#1} \fi

\bibitem[\protect\citeauthoryear{{Battaglia}
  et~al.}{2023}]{2023A&A...679A.139B}
\begin{barticle}
\bauthor{\bsnm{{Battaglia}}, \binits{A.F.}},
\bauthor{\bsnm{{Hudson}}, \binits{H.}},
\bauthor{\bsnm{{Warmuth}}, \binits{A.}},
\bauthor{\bsnm{{Collier}}, \binits{H.}},
\bauthor{\bsnm{{Jeffrey}}, \binits{N.L.S.}},
\bauthor{\bsnm{{Caspi}}, \binits{A.}},
\bauthor{\bsnm{{Dickson}}, \binits{E.C.M.}},
\bauthor{\bsnm{{Saqri}}, \binits{J.}},
\bauthor{\bsnm{{Purkhart}}, \binits{S.}},
\bauthor{\bsnm{{Veronig}}, \binits{A.M.}},
\bauthor{\bsnm{{Harra}}, \binits{L.}},
\bauthor{\bsnm{{Krucker}}, \binits{S.}}:
\byear{2023},
\batitle{{The existence of hot X-ray onsets in solar flares}}.
\bjtitle{\aap}
\bvolume{679},
\bfpage{A139}.
\doiurl{https://doi.org/10.1051/0004-6361/202347706}.
\adsurl{2023A&A...679A.139B}.
\end{barticle}
\endbibitem

\bibitem[\protect\citeauthoryear{{da Silva} et~al.}{2023}]{2023MNRAS.tmp.2291D}
\begin{botherref}
\oauthor{\bsnm{{da Silva}}, \binits{D.F.}},
\oauthor{\bsnm{{Hui}}, \binits{L.}},
\oauthor{\bsnm{{Sim{\~o}es}}, \binits{P.J.A.}},
\oauthor{\bsnm{{Valio}}, \binits{A.}},
\oauthor{\bsnm{{Costa Joaquim}}, \binits{E.R.}},
\oauthor{\bsnm{{Hudson}}, \binits{H.S.}},
\oauthor{\bsnm{{Fletcher}}, \binits{L.}},
\oauthor{\bsnm{{Hayes}}, \binits{L.A.}},
\oauthor{\bsnm{{Hannah}}, \binits{I.G.}}:
2023,
{Statistical analysis of the onset temperature of solar flares in 2010-2011}.
\textit{\mnras}.
\doiurl{https://doi.org/10.1093/mnras/stad2244}.
\adsurl{2023MNRAS.tmp.2291D}.
\end{botherref}
\endbibitem

\bibitem[\protect\citeauthoryear{{Freeland} and
  {Handy}}{1998}]{1998SoPh..182..497F}
\begin{barticle}
\bauthor{\bsnm{{Freeland}}, \binits{S.L.}},
\bauthor{\bsnm{{Handy}}, \binits{B.N.}}:
\byear{1998},
\batitle{{Data Analysis with the SolarSoft System}}.
\bjtitle{\solphys}
\bvolume{182},
\bfpage{497}.
\doiurl{https://doi.org/10.1023/A:1005038224881}.
\adsurl{1998SoPh..182..497F}.
\end{barticle}
\endbibitem

\bibitem[\protect\citeauthoryear{{Hudson} et~al.}{2021}]{2021MNRAS.501.1273H}
\begin{barticle}
\bauthor{\bsnm{{Hudson}}, \binits{H.S.}},
\bauthor{\bsnm{{Sim{\~o}es}}, \binits{P.J.A.}},
\bauthor{\bsnm{{Fletcher}}, \binits{L.}},
\bauthor{\bsnm{{Hayes}}, \binits{L.A.}},
\bauthor{\bsnm{{Hannah}}, \binits{I.G.}}:
\byear{2021},
\batitle{{Hot X-ray onsets of solar flares}}.
\bjtitle{\mnras}
\bvolume{501},
\bfpage{1273}.
\doiurl{https://doi.org/10.1093/mnras/staa3664}.
\adsurl{2021MNRAS.501.1273H}.
\end{barticle}
\endbibitem

\bibitem[\protect\citeauthoryear{{Jakiemiec}
  et~al.}{1986}]{1986AdSpR...6f.237J}
\begin{barticle}
\bauthor{\bsnm{{Jakiemiec}}, \binits{J.}},
\bauthor{\bsnm{{Sylwester}}, \binits{B.}},
\bauthor{\bsnm{{Sylwester}}, \binits{J.}},
\bauthor{\bsnm{{Mewe}}, \binits{R.}},
\bauthor{\bsnm{{Peres}}, \binits{G.}},
\bauthor{\bsnm{{Serio}}, \binits{S.}},
\bauthor{\bsnm{{Schrijver}}, \binits{J.}}:
\byear{1986},
\batitle{{Investigation of flare heating based on X-ray observations}}.
\bjtitle{Advances in Space Research}
\bvolume{6},
\bfpage{237}.
\doiurl{https://doi.org/10.1016/0273-1177(86)90151-1}.
\adsurl{1986AdSpR...6f.237J}.
\end{barticle}
\endbibitem

\bibitem[\protect\citeauthoryear{{Kane} and
  {Anderson}}{1970}]{1970ApJ...162.1003K}
\begin{barticle}
\bauthor{\bsnm{{Kane}}, \binits{S.R.}},
\bauthor{\bsnm{{Anderson}}, \binits{K.A.}}:
\byear{1970},
\batitle{{Spectral Characteristics of Impulsive Solar-Flare X-Rays $>$ 10
  KeV}}.
\bjtitle{\apj}
\bvolume{162},
\bfpage{1003}.
\doiurl{https://doi.org/10.1086/150732}.
\adsurl{1970ApJ...162.1003K}.
\end{barticle}
\endbibitem

\bibitem[\protect\citeauthoryear{{Panos}, {Kleint}, and
  {Zbinden}}{2023}]{2023A&A...671A..73P}
\begin{barticle}
\bauthor{\bsnm{{Panos}}, \binits{B.}},
\bauthor{\bsnm{{Kleint}}, \binits{L.}},
\bauthor{\bsnm{{Zbinden}}, \binits{J.}}:
\byear{2023},
\batitle{{Identifying preflare spectral features using explainable artificial
  intelligence}}.
\bjtitle{\aap}
\bvolume{671},
\bfpage{A73}.
\doiurl{https://doi.org/10.1051/0004-6361/202244835}.
\adsurl{2023A&A...671A..73P}.
\end{barticle}
\endbibitem

\end{thebibliography}

\end{article} 
\end{document}